\documentclass[10pt,letterpaper]{article}
\usepackage[top=0.85in,left=0.85in,footskip=0.75in,marginparwidth=1in]{geometry}

\usepackage[utf8]{inputenc}

\usepackage{cite}

\usepackage{nameref,hyperref}

\usepackage{microtype}
\DisableLigatures[f]{encoding = *, family = * }

\raggedright
\usepackage{changepage}
\usepackage{physics}

\usepackage[aboveskip=1pt,labelfont=bf,labelsep=period,singlelinecheck=off]{caption}

\makeatletter
\renewcommand{\@biblabel}[1]{\quad#1.}
\makeatother

\usepackage{lastpage,fancyhdr,graphicx}
\usepackage{epstopdf}
\fancyheadoffset[L]{2.25in}
\fancyfootoffset[L]{2.25in}

\usepackage{color}

\definecolor{Gray}{gray}{.25}

\usepackage{graphicx}

\usepackage{sidecap}

\usepackage[pscoord]{eso-pic}
\usepackage[fulladjust]{marginnote}
\usepackage{subcaption}
\usepackage{physics}

\reversemarginpar

\begin{document}
\vspace*{0.1in}
\centering

\begin{flushleft}
{\Large
\textbf\newline{\textit{Ab initio} quantum transport in AB-stacked bilayer penta-silicene using atomic orbitals}
}
\newline
\\
Eleni Chatzikyriakou\textsuperscript{1,*},
Padeleimon Karafiloglou\textsuperscript{2},
Joseph Kioseoglou\textsuperscript{1},
\\
\bigskip
\bf{1} Department of Physics, Aristotle University of Thessaloniki, 54124 Thessaloniki, Greece. Tel: +30 2310 996000; E-mail: elchatz@auth.gr
\\
\bf{2} Laboratory of Applied Quantum Chemistry, Department of Chemistry, POB 135, Aristotle University of Thessaloniki, 54124 Thessaloniki, Greece
\\
\bigskip
* elchatz@auth.gr

\end{flushleft}

\justify

\section*{Abstract}
The current carried by a material subject to an electric field is microscopically inhomogeneous and can be modelled using scattering theory, in which electrons undergo collisions with the microscopic objects they encounter. We herein present a methodology for parameter-free calculations of the current density from first-principles using Density Functional Theory, Wannier functions and scattering matrices. The methodology is used on free-standing AB-stacked bilayer penta-silicene. This new Si allotrope has been proposed to have a higher stability than any of its hexagonal bilayer counterparts. Furthermore, its semiconducting properties make it ideal for use in electronic components. We unveil the role of the p\textsubscript{z} orbitals in the transport through a three-dimensional quantum wire and present current density streamlines that reveal the locations of the highest charge flow. The present methodology can be expanded to accommodate many electron degrees of freedom, the application of electromagnetic fields and many other physical phenomena involved in device operation.


\section*{Introduction}

Recent advances in fabrication techniques \cite{Chen2018} and inkjet printing of 2D materials \cite{Hu2017} have brought principles of mesoscopic physics in the foreground in order to explain and predict electron device behaviour in a wide range of occasions. In resistive switching memory devices (RRAM), where a conductive filament is formed under the application of a voltage at the electrodes \cite{Trapatseli2016,Fan2017}, conductance quantization has been observed \cite{Li2015} and the SET operation has been modelled using the Landauer formula for electron tunnelling \cite{Li2017}. 'Filaments' have also been studied in relation to the atomic structure of graphene sheets theoretically \cite{Wilhelm2014} while the transition from quantum to classical regime with the use of weighted phonon self-energies has been modelled using networks of sites and the Keldysh Green's function formalism. \cite{Morr2017} These methods can serve as a valuable tool for explaining experimental topography \cite{Willke2015} and scanning tunnelling potentiometry. \cite{Lupke2017} 

Silicon has historically been the most widely used semiconducting material in the electronics industry both due to its abundance in nature and due to the low-defect interface that it forms with the insulating SiO$_2$ used in transistors and other electronic components. Hexagonal silicene\cite{Tao2015,Zhao2016} was one of the first two-dimensional sensations with a plenitude of novel properties, however, when compared to its Dirac cone counterpart, graphene, silicene has a much smaller elastic constant \cite{Sahin2009} and this poses difficulties in its manipulation in the lab. 

Other forms of silicon derivatives have been the focus of studies using systematic materials search, revealing novel allotropes with direct band-gaps \cite{chae2017} and even high mobility values \cite{zhuo2018}. Penta-silicene is a new form of monolayer silicon that has been observed recently \cite{Cerda2016,sheng2017}. Pentagons are rare atomic configurations that are less frequently studied than hexagonal structures mainly due to their difficulties imposed in their fabrication. Using first principles calculations, bilayer penta-silicene, whose layers are stacked in 90$^o$ angle between them has been found to be more stable than the most stable form of bilayer hexagonal silicene \cite{Aierken2016}. It is made up of pentagonal rings of Si atoms, similar to penta-graphene \cite{Logoteta2014} and other two-dimensional penta-structures \cite{Liu2016_2}. Twisting angles between the layers of few-layer materials has been proved to be an efficient method for tuning their properties \cite{Rivera2015,Chiu2014}. Contrary to the bilayer form with no twisting angle, the AB stacking configuration of penta-silicene has semiconducting properties. This makes it ideal for many different configurations of electronic devices, stemming from the recent surge in heterostructure fabrication \cite{Liu2016}.

In the structures that we examine herein, parts of the system lose their symmetry in favour of stability. Situations such as this, call for the more intuitive chemical description of the system in terms of its valence bonding configuration and this can be very well described with the use of maximally localized Wannier functions (WFs). The latter, acting as alternative representations of Bloch orbitals, have recently gained popularity for examining a plethora of mesoscopic phenomena due to advancements in localization techniques \cite{Souza2001,Thygesen2005} but also when used as tools for high-throughput screening of topological materials \cite{Gresch2017}. Even more interesting for this study, is the possibility to derive model Tight-Binding Hamiltonians in the WF basis, with specific orbitals involved (i.e. excluding higher conductions bands), arbitrary number of neighbor terms and cut-off values to adjacent cells, and therefore increased accuracy and reduced computational cost, suited to multi-scale modelling of 2D semiconductor transistors \cite{Pizzi2016, Iannaccone2018}.

Therefore, compared to previous work using Kohn-Sham states and mesoscopic transport using Green's functions \cite{Walz2015}, the use of Wannier functions provides more computational advantages. Finally, the method of scattering matrices is both fast and stable \cite{Groth2014} and can be expanded to include, among others, spin and other degrees of freedom \cite{Schaibley2016}, electromagnetic fields and time-dependent effects \cite{Weston2016}.

\begin{figure}[h!]
\centering
  \includegraphics[height=4cm]{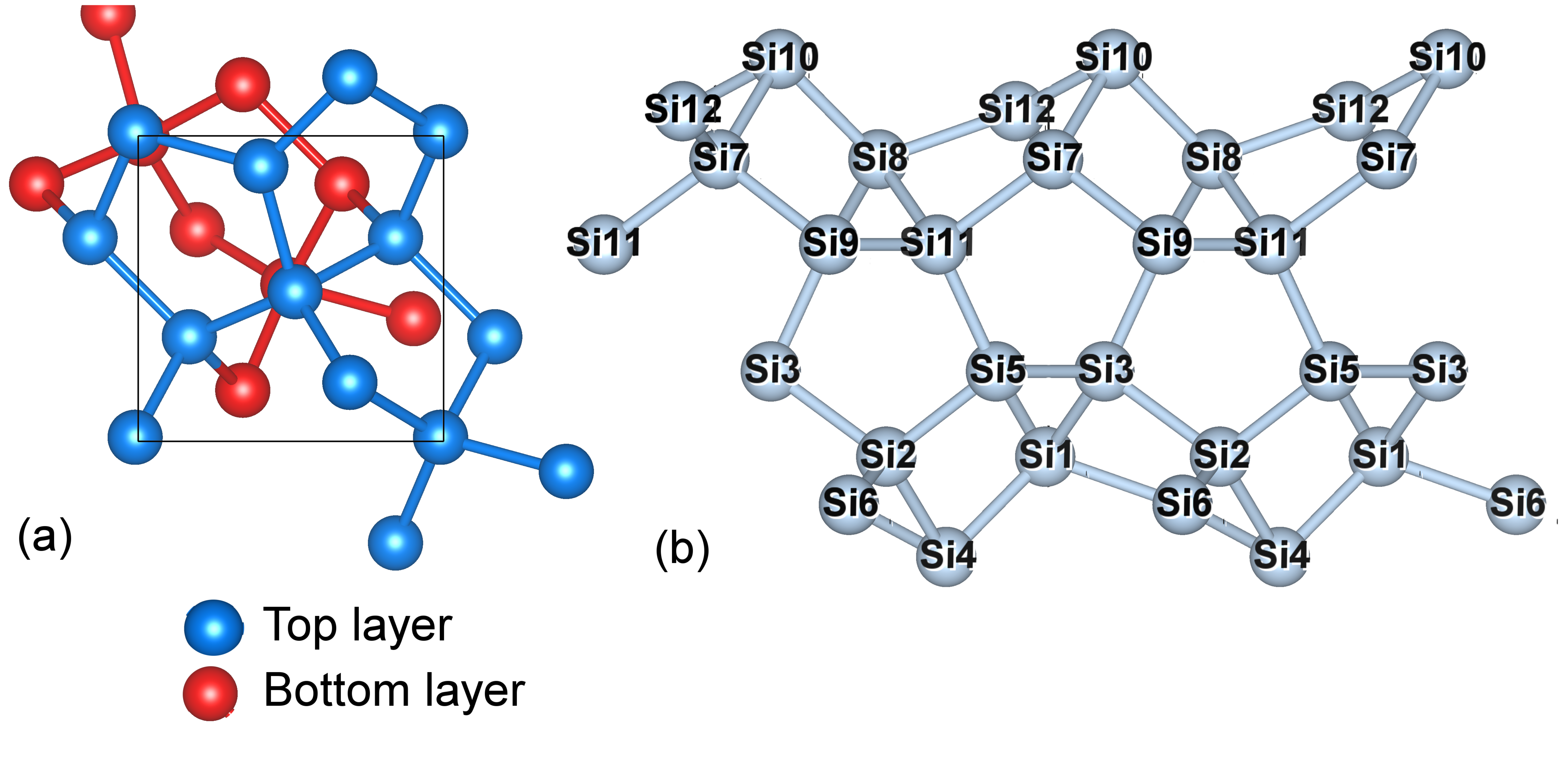}
  \caption{(a) Bilayer form of penta-silicene with AB stacking configuration. Top and bottom layer atoms are shown with different color (b) Lateral view of the structure showing also the position of its atom in the crystal.}
  \label{fgr:BiPentaSi_ABdr}
\end{figure}

\section{Results \& Discussion}

\begin{figure}[t]
\centering
  \includegraphics[height=6.2cm]{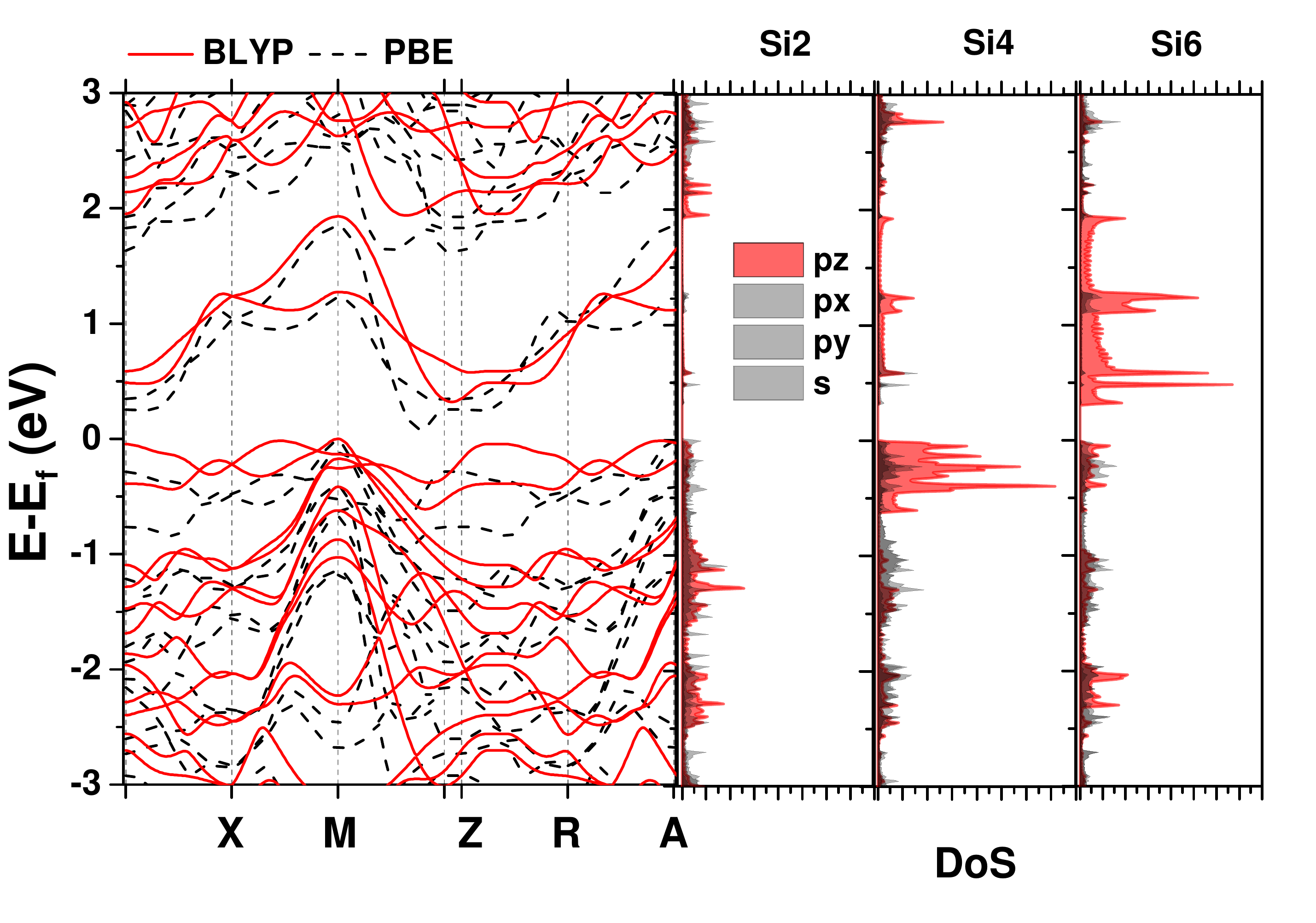}
  \caption{Bandstructure diagram of AB stacked bilayer penta-silicene and orbital projected DoS of atoms 2,4 and 6.}
  \label{fgr:Band_Orbitals}
\end{figure}

Figure \ref{fgr:BiPentaSi_ABdr}a shows the unit cell used in the calculation. It contains two layers of Silicon with a total of 12 atoms (6 at the top and 6 at the bottom layer) \cite{Aierken2016}. Each atom is labelled with a number (Si1-Si12) as shown in Figure \ref{fgr:BiPentaSi_ABdr}b. The computational details are shown in the Methods section. The well-known bandgap-problem of DFT becomes relevant when considering device characteristics, as current transport is mainly performed by the states around it. This can be overcome using hybrid exchange correlation functionals, Hartree-Fock methods or the GW approximation.\cite{Iannaccone2013} We have compared band structure results for two different exchange-correlation functionals (Fig. \ref{fgr:Band_Orbitals}). The calculated bandgap is indirect with 0.321 eV for BLYP and 0.107 eV for the PBE functional, while the minimum of the conduction band is located between $\Gamma$Z in the first case and between $\Gamma$M in the second. The maximum of the valence band in located at M in both cases. The PBE functional was used further in this work, however, we can arrive at similar conclusions using any of the aforementioned methods.

Each layer of the structure has tetragonal symmetry and each sublattice is twisted 90$^o$ to the other one as shown in Figure \ref{fgr:BiPentaSi_ABdr}a. Aierken at al. \cite {Aierken2016} showed that the structure was stable when the electrons on each of the surfaces of its free-standing form lose their symmetry (Figure \ref{fgr:BiPentaSi_ABdr}b). It is natural that when the material is fabricated on various substrates, depending on the method used and substrate material, bonding and hybridization will be adjusted. Cerdá et al. showed that the pentagons that are formed on Ag(110) host both sp\textsuperscript{2} and sp\textsuperscript{3} bonded Si atoms, in which the latter also bond with Ag substrate atoms.\cite{Cerda2016} For hexagonal silicene nanoribbons, there has been increased interest in their edge states as doping and hydrogenation has shown to change their electronic and magnetic properties \cite{Ding2013,Ding2014}. 

Using orbital projected calculations with DFT, the Lowdin charges \cite{Lowdin1950} of the atoms in the structure were derived (Table \ref{tbl:Lowdin}). All atoms, except for two at each surface (Si4,6,10 and 12) were found to have total charge approximately 4, of which roughly 30\% resides in s states and 70\% in p states. Then, on each of the surfaces, the Si6 and Si12 atoms lose part their charge to the Si4 and Si10 atoms equivalently, whose hybrid orbitals acquire 35\% s and 65\% p character.   

\begin{table}[h]
\small
  \caption{\ Total charges and charges in s and p states (in number of electrons) of the Si atoms. Note that the calculations performed are spin-unpolarized.}
  \label{tbl:Lowdin}
  \begin{tabular*}{0.48\textwidth}{@{\extracolsep{\fill}}llll}
    \hline
    Atom No. & Total charge (e) & s (e) & p (e) \\
    \hline
    Si1-3,5,7-9,11 & 3.95 & 1.16 & 2.78 \\
    Si4,10 & 4.12 & 1.45 & 2.67 \\
    Si6,12 & 3.74 & 1.29 & 2.44 \\
    \hline
  \end{tabular*}
\end{table}

Orbital projected Density of States (DoS) for three representative atoms is plotted next to the band structure in Figure \ref{fgr:Band_Orbitals}. Most of the atoms have contributions equivalent to that of Si2, except for the two surface atoms that were described previously, for which their p\textsubscript{z} orbitals show increased contribution around the Fermi level. The bottom of the conduction band is formed by the p\textsubscript{z} orbitals of Si6 and Si12, while the top of the valence band is formed by the p\textsubscript{z} orbitals of the Si4 and Si10.

\begin{figure}[t!]
  \centering
  \includegraphics[height=5.5cm]{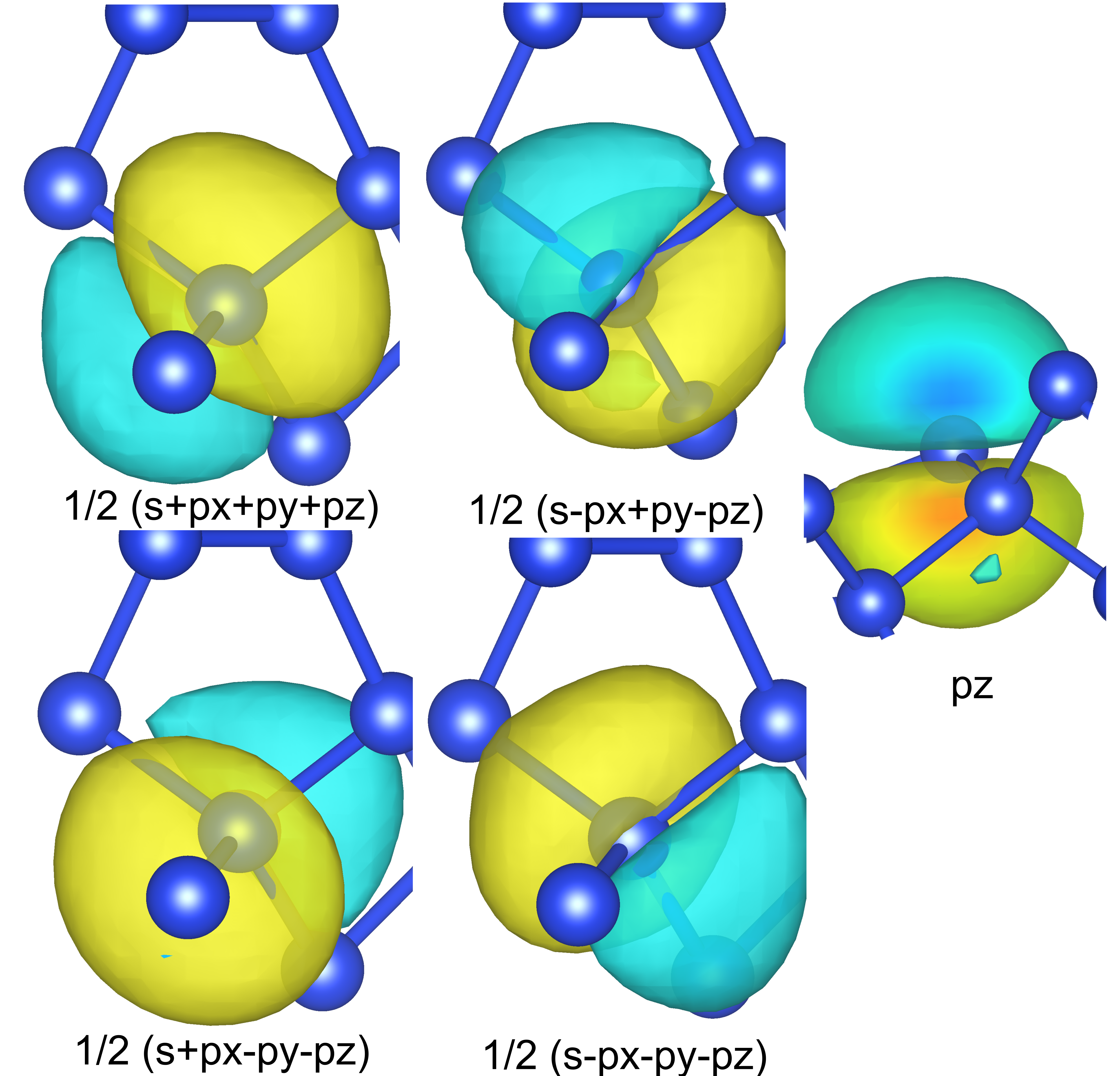}
  \caption{WFs of two atoms. (a)-(d) show the four orientations of the orbitals in the sp\textsuperscript{3} hybridized atom (e) shows the p\textsubscript{z} orbital of an sp\textsuperscript{2} hybridized atom.}
  \label{fgr:WannierOrbitals}
\end{figure}

Wannier90 \cite{Wannier90} was used for the Wannierization. In order to derive the effective Hamiltonian from the atomic orbitals, 48 projections were used with atom-centered orbitals, namely, 8 sp\textsuperscript{3}, 4 sp\textsuperscript{2} and 4 p\textsubscript{z}, as they were found in the orbital-projected DFT calculations. In Figure \ref{fgr:WannierOrbitals}, the plots of 5 WFs are given. The results correctly reflect the sp\textsuperscript{3} nature of the equivalent atoms and the remaining p\textsubscript{z} orbitals of the distorted symmetry atoms.

\begin{figure}[b]
\centering
  \includegraphics[height=3.2cm]{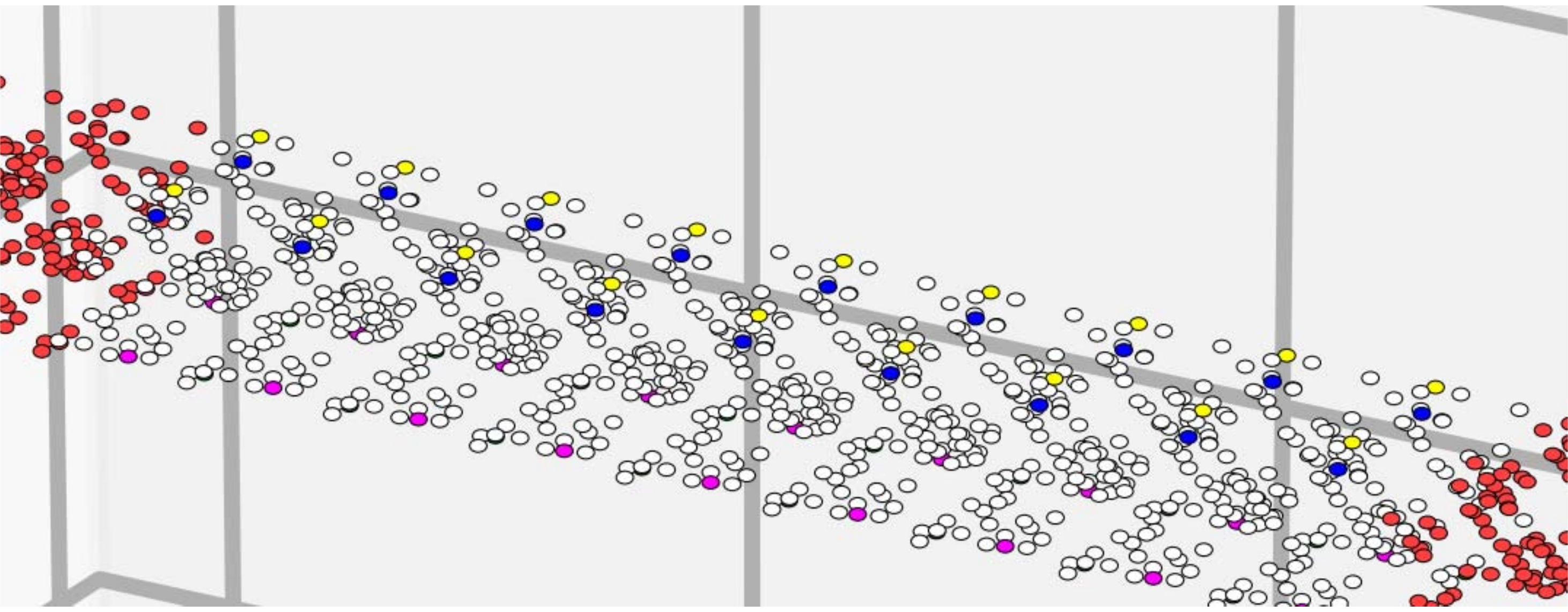}
  \caption{Kwant graph showing the sites of a 2x9 WxL quantum wire with magenta, green, blue and yellow showing the p\textsubscript{z} orbitals of Si4, Si6, Si10 and Si12 respectively. }
  \label{fgr:system_pz}
\end{figure}

\begin{figure}[h]
\centering
  \includegraphics[height=6.9cm]{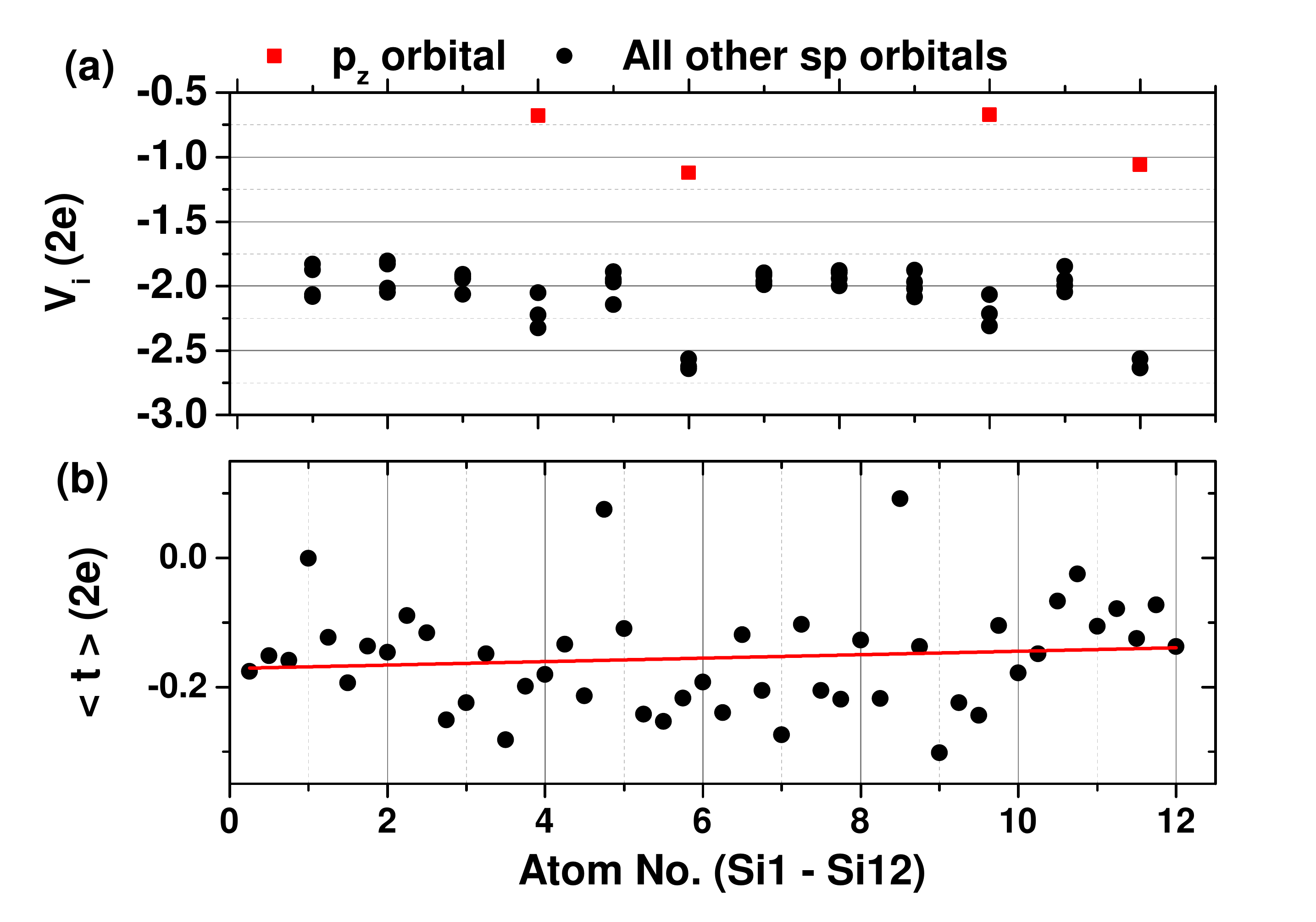}
  \caption{(a) On-site energies of the orbitals of each atom and (b) Average value of the hopping integrals showing also a slight increase towards the top orbitals of the system.}
  \label{fgr:onsite_hopping}
\end{figure}

Transport calculations on a free-standing quantum wire (Fig. \ref{fgr:system_pz}) with a scattering region of dimensions 9x2x1 unit cells were performed using Kwant \cite{Groth2014}. The length expands in the $k_x$ direction. The size of the leads is 3x2x1 with 1D translational symmetry in the directions away from the scattering region. Both the scattering region and the leads are of the same material type, as this allows us to concentrate solely on its properties. A generalization of this can include leads of different material type, or a scattering region with an interface between two different materials. The solution of the Schroedinger equation in the system now corresponds to the plane wave nature of the electrons and not the Bloch waves used in DFT. The Hamiltonian of the system is given by,

\begin{equation} \label{eq1}
\hat{H}=\sum_{i}V_{i}c_i^{\dagger}c_i+t\sum_{<i,j>}\left(c_i^{\dagger}c_j+H.c.\right)
\end{equation}

where i and j denote Wannier sites with $V_{i}$=$<\phi_i|H|\phi_i>$ the diagonal elements of the Hamiltonian matrix derived after the Wannierization procedure, $c_i$ and $c_i^{\dagger}$ denote the electron creation and annihilation operators respectively, and t=$<\phi_i|H|\phi_j>$ the off-diagonal matrix elements equivalently. The on-site energies $V_i$ correspond to those of the 48 WFs of the penta-Si atomic orbitals and the hopping integrals t represent the possibility for an electron to jump from state $\phi_i$ to state $\phi_j$. The whole system is Hermitian. Expanding to further degrees of freedom could be done by representing one site with one atom in the system, and each orbital being an element in the on-site Hamiltonian matrix of the atom. The on-site and hopping integral energies were extracted using TBModels \cite{greschTBmodelsDocumentation,Gresch2017}. Spin can also be added by solving the spin-polarized Kohn-Sham equations and then separating the up and down spin components from the Wannier90 results \cite{Marzari2012}. In the case examined here, the calculations are closed-shell, and therefore, we include a factor of two in all further units that include electron charge (e). 

The on-site energies of the orbitals are given in Figure \ref{fgr:onsite_hopping}(a). Shown in red are the on-site energies of the p\textsubscript{z} orbitals of the equivalent atoms, whose energies are higher than the rest of the atomic sp orbitals. Figure \ref{fgr:onsite_hopping}(b) shows the average value of the hopping integrals for each orbital in the unit cell. Both the on-site energies and the hopping integrals show a slight increase in the atoms of the upper layer of the system. The asymmetry of the final Tight Binding Hamiltonian is a problem that arises after the disentanglement procedure and which has only recently been addressed. \cite{Gresch2018}

\begin{figure}[h]
\centering
  \includegraphics[height=7cm]{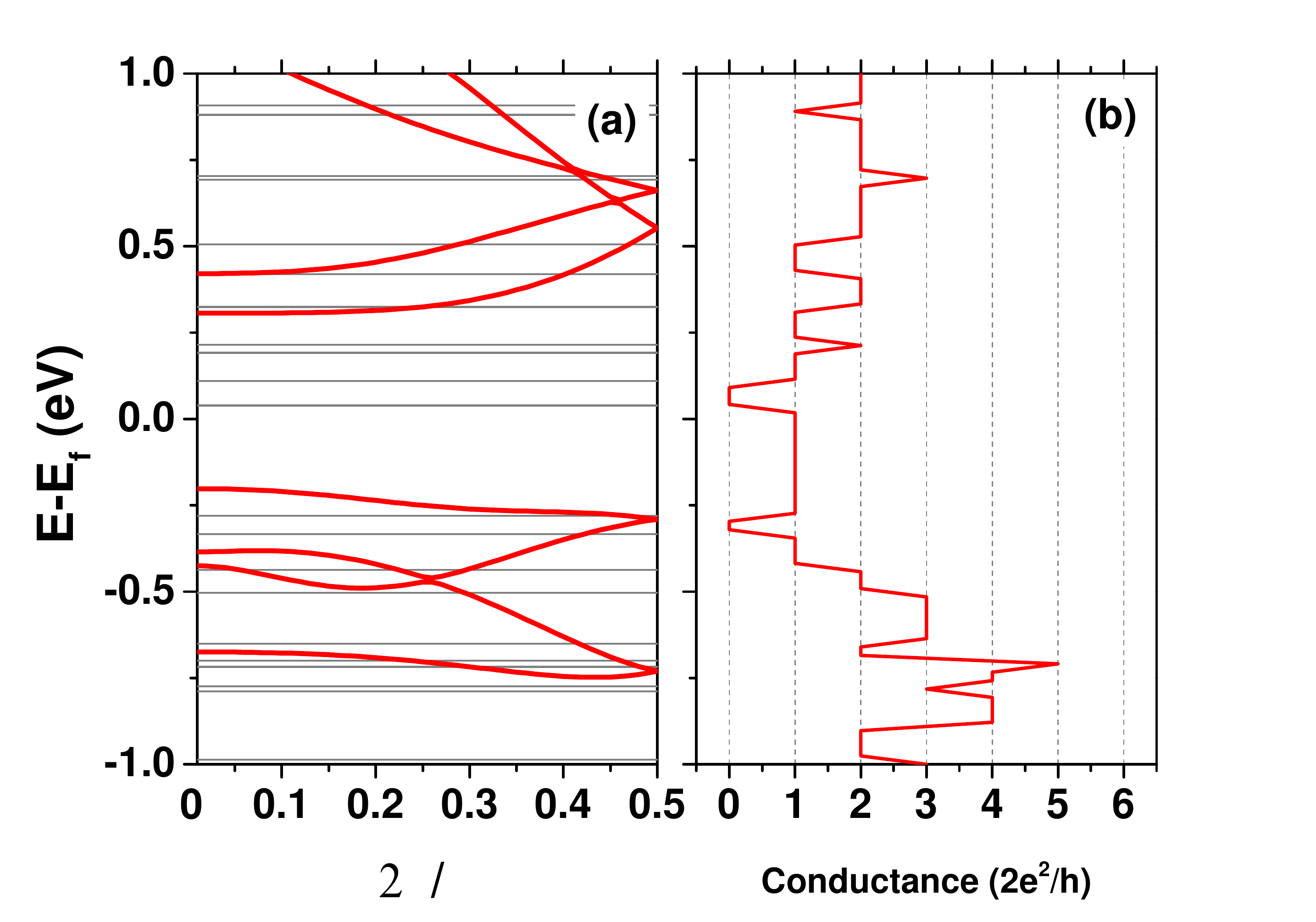}
  \caption{(a) Eigenenergies of the Hamiltonian of one unit in the lead, in the k$_x$ direction, with 3D translational symmetry (red lines) and similarly with 1D translational symmetry for 3x2 unit cells (black line) (b) Quantum conductance of the wire in the same direction.}
  \label{fgr:Cond_all}
\end{figure}

The Tight Binding (TB) Hamiltonian matrix resulting from the Wannierization is cut-off at three nearest neighbour hoppings. In this context, the nearest neighbours are the hoppings that lie within the home unit cell, second nearest neighbours are the hoppings to adjacent cells in all directions etc. The maximum order of the nearest neighbour interactions is dependent on the k-points used in the Monkhorst-pack mesh during the DFT calculations, as they define the number of periodic images that will appear when the system is translated in real space during Wannierization. As the number of nearest neighbours increases, a large number of neighbour interactions can significantly increase computation time both for creating the system and solving it. However, in most cases, three nearest neighbour interactions provide sufficient accuracy (see Suppl. information).

The conductance of the wire is calculated from the scattering matrix of the system \cite{Groth2014}. The dispersion relations within the leads that have 1D translational symmetry are shown together with the conductance plots in Figure \ref{fgr:Cond_all}. The energies of the modes take on constant values in each direction due to the lack of a 3D translational symmetry. The magnitude of the conductance reflects the probability of transmission of the modes at each electrochemical potential difference between the leads. A value of two shows the there are two modes propagating in the wire at this energy etc.

\begin{figure}[t]
\centering
  \includegraphics[height=6.5cm]{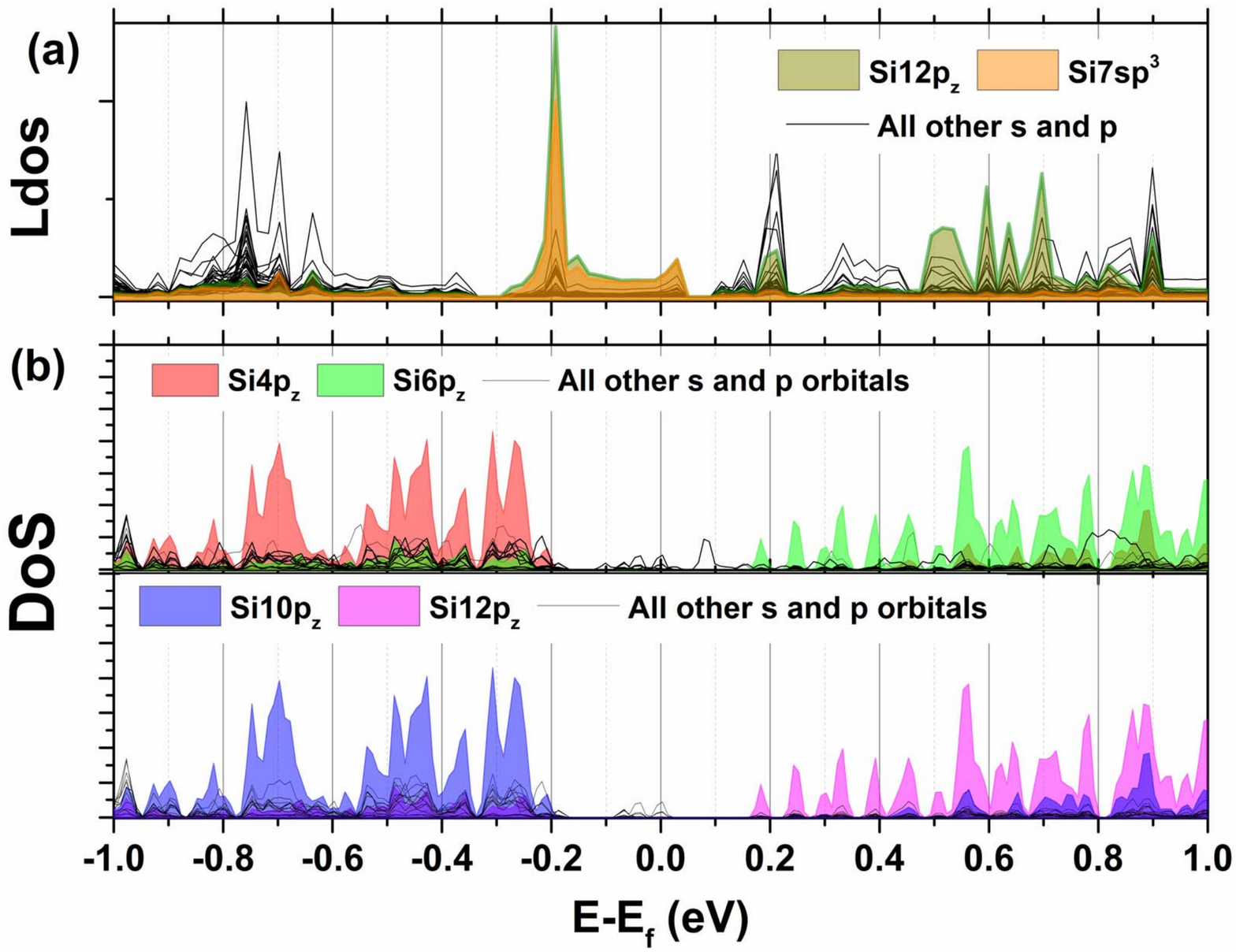}
  \caption{(a) Local density of states for each orbital type in the quantum wire (b) Orbital-projected DoS derived from DFT calculations using the PBE functional.}
  \label{fgr:Density_DoS}
\end{figure}

\begin{figure*}[h]
\centering
\begin{subfigure}{.60\textwidth}
\centering
  \includegraphics[height=7.0cm]{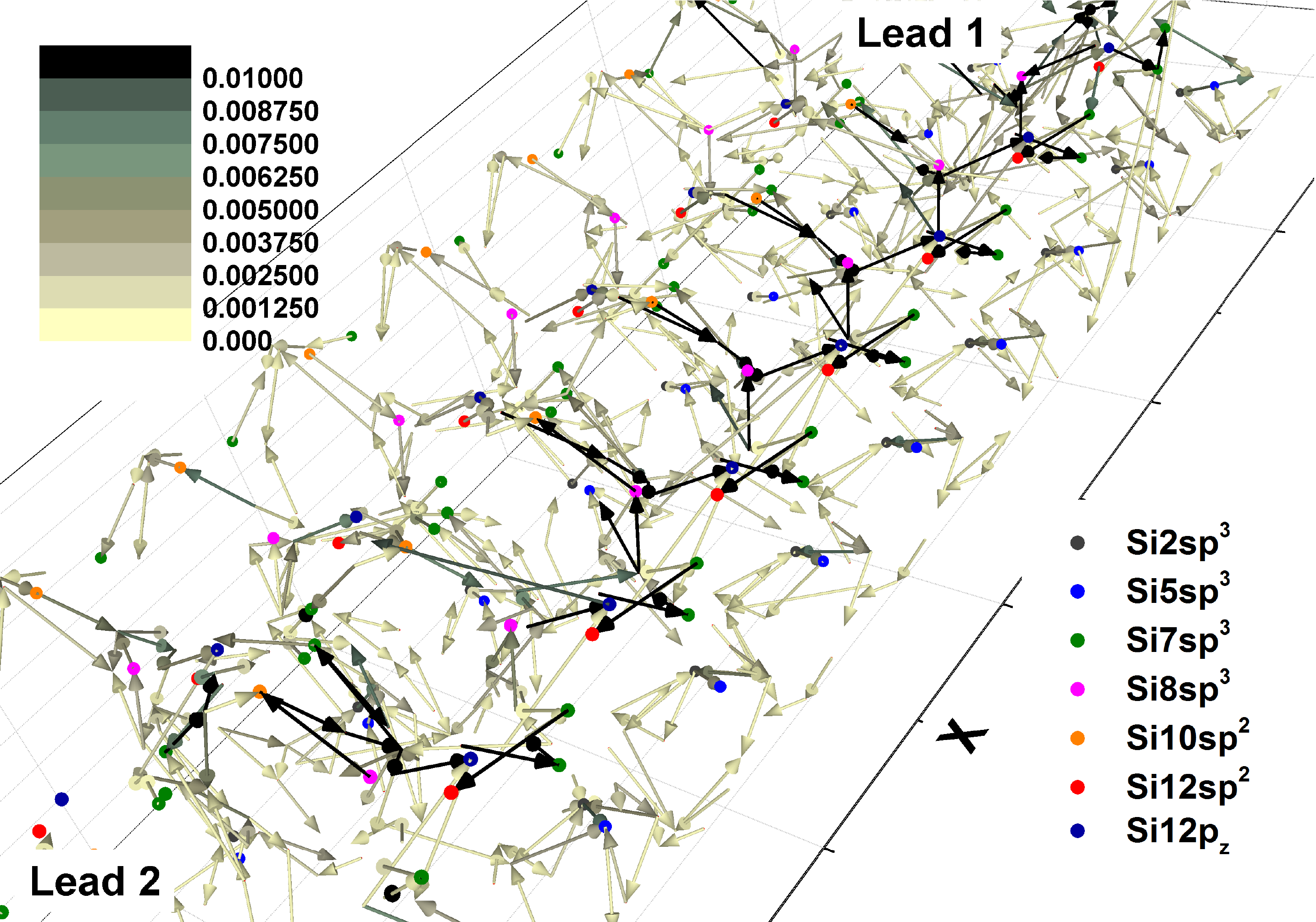}
  \caption{}
  \label{fgr:Density_current_vector}
  \end{subfigure}
  \begin{subfigure}{.30\textwidth}
\centering
  \includegraphics[height=8cm]{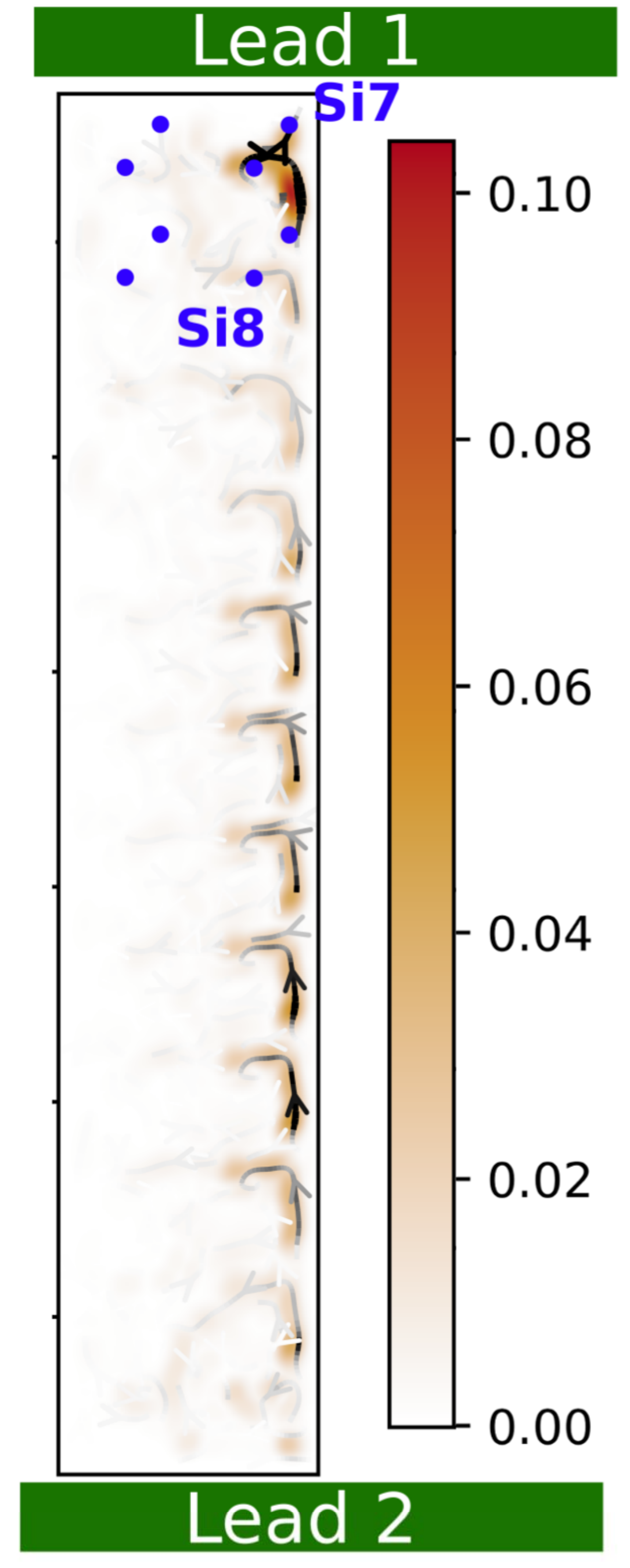}
  \caption{}
  \label{fgr:Streamlines_current}
\end{subfigure}
\caption{(a) Vector plot showing the location of the hopping with the highest current density for each site. The colour scale shows the magnitude of the expectation value (2e/h). The arrowhead shows the direction of the current at this hopping (b) Interpolated current density and streamlines at two cuts perpendicular to $k_z$ at the vicinity of the highest concentration of local current density. Transport is in the $k_x$ direction (Lead 1 towards Lead 2). The locations of the atoms that exist at this z cut (with an approximate displacement of 1 \r{A}) are marked in blue.}
\label{fig:current}
\end{figure*}

Local charge and current densities are accessible through solving for individual or pairs of sites in the system, giving access to computations of many properties that exist at surfaces and interfaces. Figure \ref{fgr:Density_DoS}(a) shows the averaged local density of states from each orbital type which is taken from the expectation value of the local density operator in Kwant using,
 
\begin{equation} \label{eq4}
\frac{\sum_k{\sum_n{\bra{\psi}Q_n\ket{\psi}}}}{k}
\end{equation}

where $\psi$ is any propagating wavefunction in the scattering region at each energy, n runs over all expectation values resulting from all incoming and outgoing wavefunctions and k runs all sites representing its orbital type in the system.   

At E\textsubscript{f}-0.19 eV, an increased electron density is observed at Si7 sp\textsuperscript{3} and Si12 p\textsubscript{z} orbitals.  To explain this, the expectation value of the current density at this energy was also extracted for the propagating wavefunctions originating from lead 1 (left lead in Fig. \ref{fgr:system_pz}). For a hopping from site l to site k, this is calculated from,

\begin{equation} \label{eq6}
\sum_n{J_{k,l}} = i\sum_n{\left[\bra{\psi_{l}}\left(H_{kl}\right)^{\dagger}\ket{\psi_k} - \bra{\psi_k}H_{kl}\ket{\psi_l}\right]}
\end{equation}

where n runs over all aforementioned propagating wavefunctions. One of the many advantages of using this method is that it gives us access to three-dimensional quantities. To visualize the results we have used both a vector plot and a 2D cut of the scattering region. Figure \ref{fgr:Density_current_vector} shows the magnitude and direction of the highest hop for each site in the system. For a more clear view, figure \ref{fgr:Streamlines_current} shows a 2D cut at the location of the highest current density in the system and the atoms that are located at its vicinity for the first four unit cells. It is seen more clearly that transport occurs in the left most side of the wire.

In the absence of any potential or magnetic field defined explicitly over the 3D region, the distribution of the current density is dictated by the confinement effects in the x,y and z directions, \cite{Wilhelm2014} that is by the solution of the Schroedinger equation with the translation operator in the leads defining the energies of the propagating modes. Finally, the on-site energies as well as the hopping integral values define the electrostatic landscape for the electrons to find their route through the wire (Fig. \ref{fgr:onsite_hopping}). \footnote[4]{While writing this manuscript the authors were made aware of another work that adds to the Wannierization procedure by creating symmetrized Tight Binding models. \cite{Gresch2018} This is expected to significantly increase the accuracy of the ideal ground state model Hamiltonian.} 

Taking into account collective effects from many orbitals, the effects of impurities and screening or the formation of dipoles at interfaces \cite{salaoru2017} can furthermore be examined opening the road to versatile computational microscopic and topographic studies, while phonon effects can be added as site self-energies using standard procedures. \cite{Morr2017} Progress in obtaining fast and accurate model Tight Binding model Hamiltonians is currently flourishing \cite{PythTBDocumentation} and is expected to lead to significant advances in predictions of device characteristics.

\section{Conclusions}

An \textit{ab initio} multi-scale simulation approach has been presented for the calculation of the current density between adjacent atomic positions in a material using an effective Hamiltonian derived from Wannier functions. The methodology has been applied to a newly predicted material by the name bilayer penta-silicene, where we have observed an increased concentration of charge at specific orbitals in a free-standing quantum wire. This was found to be consistent with the expectation values of the DC local current between its atomic orbitals, which revealed the locations of the highest flow of charge.
 
This methodology presents many advantages for the examination of electron device operation as it allows many microscopic details of the quantum transport to be revealed using realistic values and the addition of disorder, spin, phonon and finite-temperature effects. Bias calculations are also possible by summing the propagating modes around the fermi level in the scattering region.\cite{FoaTorres2016}

\section*{Conflicts of interest}
There are no conflicts to declare.

\section*{Acknowledgements}
This work was supported by computational resources granted from the Greek Research \& Technology Network (GRNET) in
the National HPC facility \lq ARIS\rq \ under the project AMONADE (ID pr004002) and also used the European Grid Infrastructure (EGI) through the National Grid Infrastructures NGI\_GRNET, HellasGRID as part of the SEE Virtual Organisation. Data presented in this paper are available at DOI: 10.5281/zenodo.1438840.

\section*{Methods}

\subsection*{DFT}
Density functional theory calculations were performed using the Quantum Espresso package \cite{QE-2017}, norm-conserving Goedecker/Hartwigsen/Hutter/Teter pseudopotential with BLYP\footnote[4]{\href{http://www.quantum-espresso.org/wp-content/uploads/upf\_files/Si.blyp-hgh.UPF}{http://www.quantum-espresso.org/wp-content/uploads/upf\_files/Si.blyp-hgh.UPF}} and PBE\footnote[5]{\href{http://www.quantum-espresso.org/wp-content/uploads/upf\_files/Si.pbe-hgh.UPF}{http://www.quantum-espresso.org/wp-content/uploads/upf\_files/Si.pbe-hgh.UPF}} exchange correlation functionals. 96 bands were included in the calculation for the 12 atoms of the unit cell, each with 4 valence electrons (3s2 3p2). The plane wave cut-off energy was set to 37 Ry and a Monkhorst-pack k-point mesh of 9x9x1 was used for the relaxation and the band structure calculations. These settings resulted in a lattice constant of 5.29 \r{A} for the case of the BLYP and 5.21 \r{A} for PBE.

\subsection*{Wannier calculations}

Wannier90 \cite{Wannier90} calculations were performed with PBE functional DFT results given as input. A frozen window that included the 26 valence and 22 conduction bands was defined. The tolerance for the gauge invariant term of the spread of the WF was set to the really low value of 10$^{-7}$ \r{A}$^2$ as the bands towards higher energies are highly entangled. The density of the k-point mesh was explicitly optimized to 15x15x1 k-points. Imaginary/Real ratios of the order 10$^{-3}$--10$^{-4}$ were achieved for the WFs. 


\bibliography{library}

\bibliographystyle{unsrt}

\newpage

\section*{Supplementary information}

\subsection*{Wannier functions}

\begin{figure}[h]
\centering
  \includegraphics[height=6.0cm]{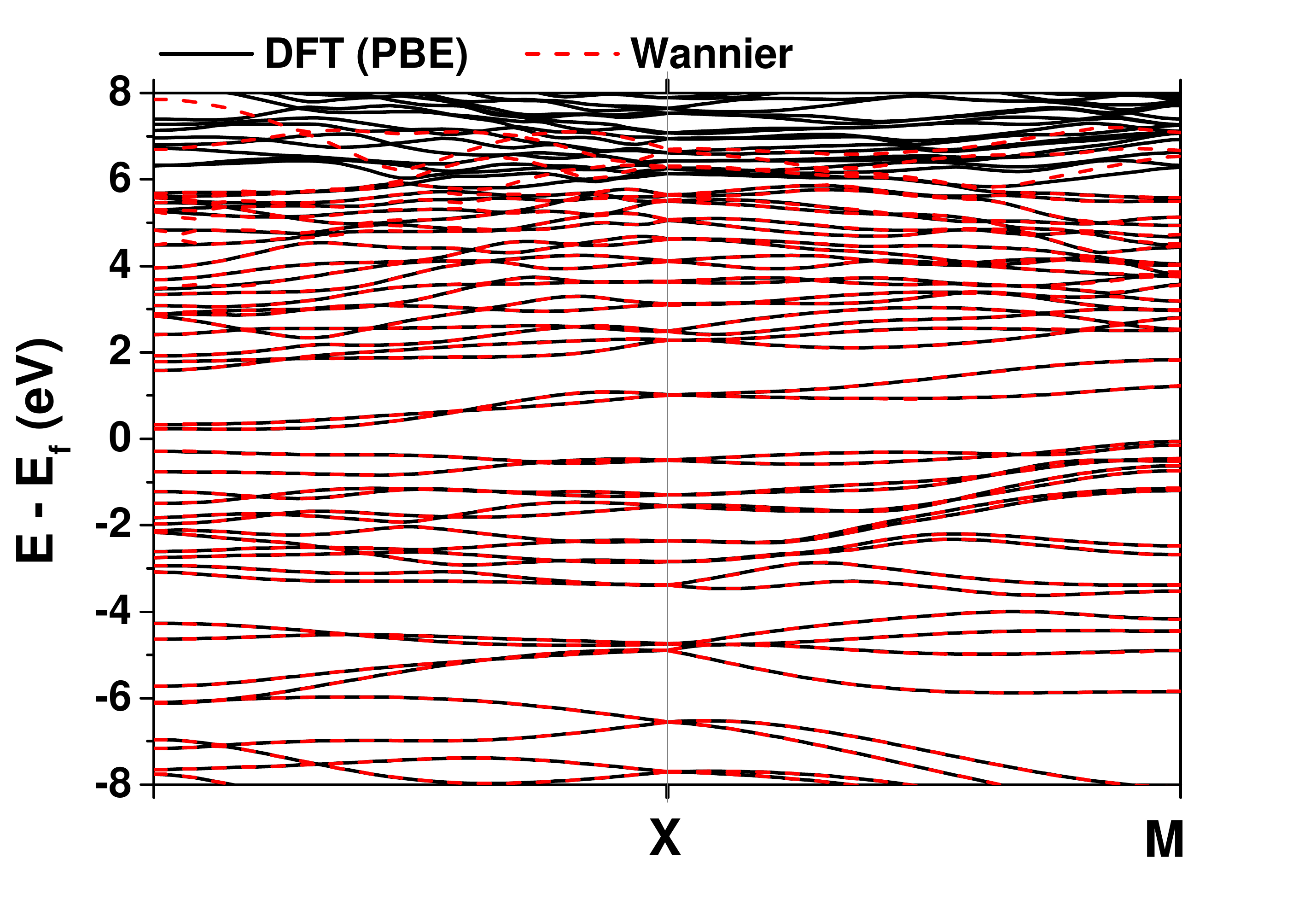}
  \caption{Comparison between the band structure results from Wannier functions and DFT calculations with PBE exchange correlation functional.}
  \label{fgr:Wannier_DFT}
\end{figure}

The convergence of the Wannier functions was checked both from their imaginary parts and from plotting the bandstructure. Figure \ref{fgr:Wannier_DFT} shows a comparison of the bandstructure in a portion of the reciprocal space from Wannier and DFT results. Higher in the conduction band, the highly entangled eigenstates reproduce poorly, however, for transport simulations, we are mainly interested in the states around the bandgap. Even in this case, where most of the orbitals acquire a less localized sp nature, the system was efficiently simulated.

\subsection*{Band structure from Tight Binding model}

\begin{figure}[h]
\centering
  \includegraphics[height=7.0cm]{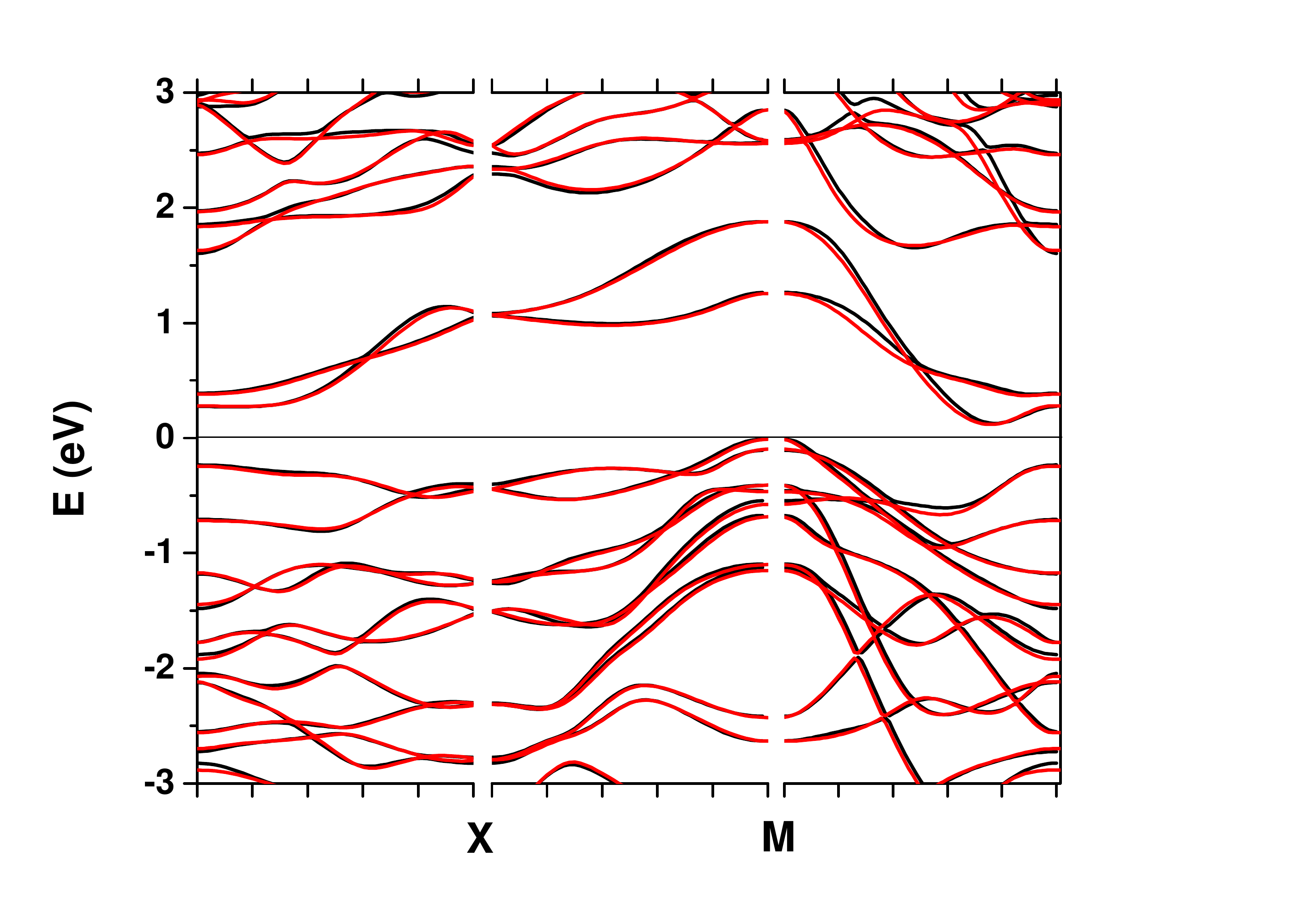}
  \caption{Dispersion relations in Kwant (black lines) and band structure from DFT code with PBE functional (red lines).}
  \label{fgr:Bands_kwant_DFT}
\end{figure}

The band structure was checked for its correctness in the Kwant code \cite{Groth2014}. A 1x1 system with three nearest neighbours was used in Kwant with 3D translational symmetry, where the dispersion relations were derived for the high symmetry points in the Brillouin zone (fig. \ref{fgr:Bands_kwant_DFT}). 

\subsection*{Quantum conductance}

\begin{figure}[b!]
\centering
  \includegraphics[height=6.9cm]{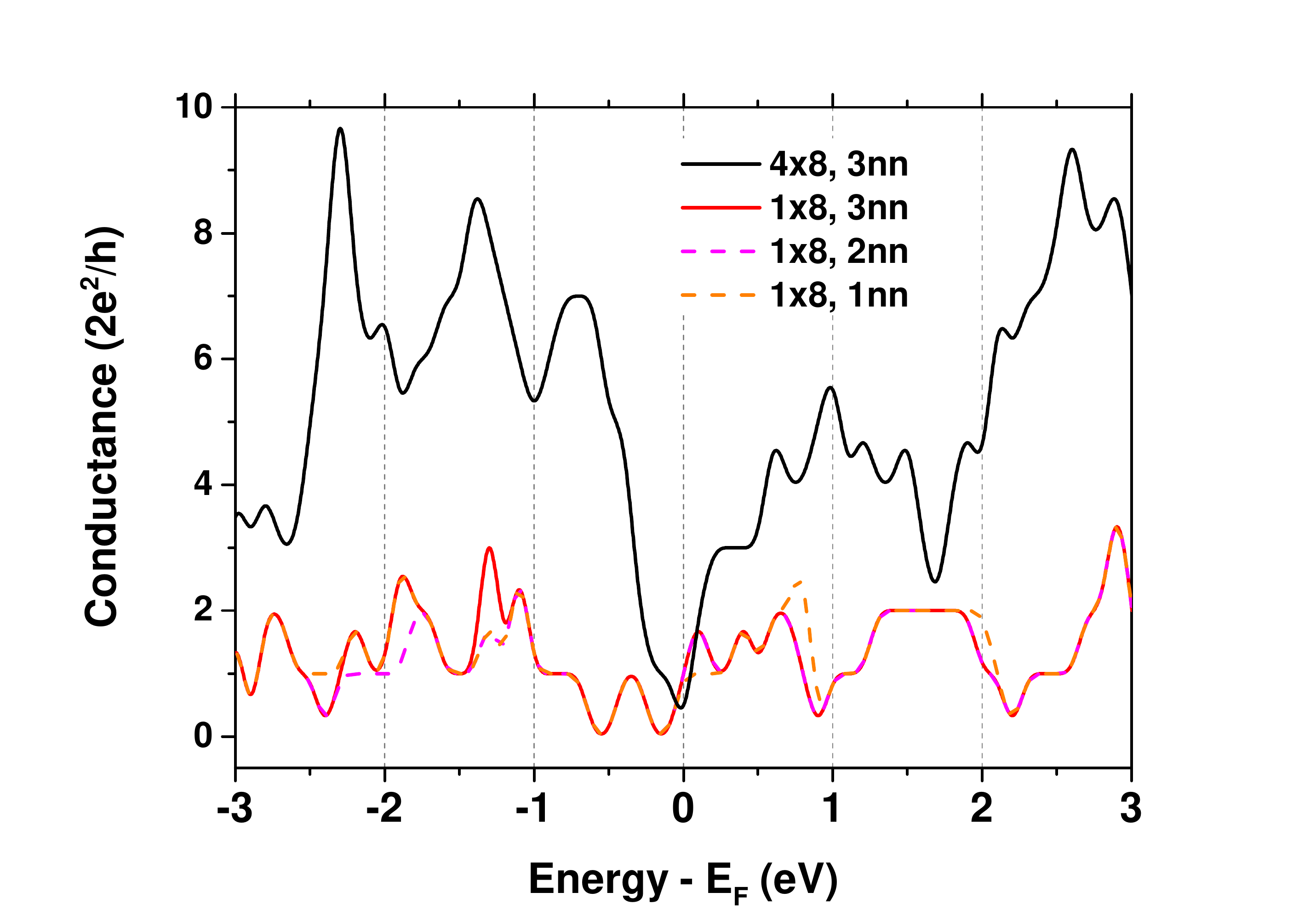}
  \caption{Conductance results for a scattering region of 1x8 and 4x8 W/L and nearest neighbour hoppings of order (nn) as indicated - 1 being next nearest neighbours. The energy step is 0.1 eV. The length extends in the $k_x$ direction.}
  \label{fgr:Conductance}
\end{figure}

The accuracy of quantum transport simulation increases with the order of the nearest neighbor (nn) hoppings. Conductance calculation results with different nn hoppings are shown in Figure \ref{fgr:Conductance} with 0 being nearest neighbour, 1 next-nearest neighbour and so on. The conductance is also equal to \cite{Groth2014},

\begin{equation} \label{eq2}
G_{ab}=\frac{2e^2}{h}MT=\frac{2e^2}{h}\sum_{n\in a,m\in b}\left|S_{nm}\right|^2
\end{equation}

We observe differences in the conductance of the medium when allowing hoppings to more distant neighbours which is attributed to the differences in the scattering probability that occurs due to the changes in the eigenenergies of the Hamiltonian in the leads. Increasing the width of the wire amounts to an increase in the magnitude of its conductance in discrete steps, as more modes are propagating in the wire, while with a sufficient increase in the number of atoms in the system, the Ohmic behaviour is restored \cite{datta_1995},

\begin{equation} \label{eq3}
G_{ab}=\sigma\frac{W}{L}
\end{equation}

where $\sigma$ is the conductivity of the material, W is the width and L is the length of the semiconductor region. 

For calculating current-voltage characteristics of the wire, summing over all propagating modes close to the fermi level would be needed. In this case, finite temperature effects and any non-linearities will properly need to be addressed.

\end{document}